# Analytical modeling of demagnetizing effect in magnetoelectric Ferrite/PZT/Ferrite trilayers taking into account a mechanical coupling.

V. Loyau, A. Aubert, M. LoBue, and F. Mazaleyrat

*SATIE UMR 8029 CNRS, ENS Cachan, Université Paris-Saclay, 61, avenue du président Wilson, 94235 Cachan Cedex, France.*

*Abstract.* In this paper, we investigate the demagnetizing effect in ferrite/PZT/ferrite magnetoelectric (ME) trilayer composites consisting of commercial PZT discs bonded by epoxy layers to Ni-Co-Zn ferrite discs made by a reactive Spark Plasma Sintering (SPS) technique. ME voltage coefficients (transversal mode) were measured on ferrite/PZT/ferrite trilayer ME samples with different thicknesses or phase volume ratio in order to highlight the influence of the magnetic field penetration governed by these geometrical parameters. Experimental ME coefficients and voltages were compared to analytical calculations using a quasi-static model. Theoretical demagnetizing factors of two magnetic discs that interact together in parallel magnetic structures were derived from an analytical calculation based on a superposition method. These factors were introduced in ME voltage calculations which take account of the demagnetizing effect. To fit the experimental results, a mechanical coupling factor was also introduced in the theoretical formula. This reflects the differential strain that exists in the ferrite and PZT layers due to shear effects near the edge of the ME samples and within the bonding epoxy layers. From this study, an optimization in magnitude of the ME voltage is obtained. Lastly, an analytical calculation of demagnetizing effect was conducted for layered ME composites containing higher numbers of alternated layers ($n \geq 5$). The advantage of such a structure is then discussed.

## I. Introduction.

Magnetoelectric (ME) composites using the product-property concept are particularly suitable for smart sensors fabrication (e.g. magnetic field or current sensors[1-5]). The product-property effect is obtained when piezoelectric and magnetostrictive phases are mechanically coupled to each other. At the present date, layered ME composites have high interest because they produce the best ME performances. Bilayers of piezoelectric and magnetostrictive materials are the simplest layered composites but these structures exhibit low ME effects. In order to achieve high ME responses, some authors[6,7] have focused their studies on co-sintered ME samples containing a high number of alternated PZT/ferrite thin layers. Among the different structures of layered composites, the trilayer, consisting of a piezoelectric layer sandwiched between two magnetostrictive layers, achieves a good balance between ease of fabrication and performances[8,9]. In a recent paper[4], we have shown that a piezoelectric layer stressed on its two faces by two ferrite layers (ME trilayer) has much better mechanical coupling in comparison with a bilayer configuration where the piezoelectric layer is stressed only on one face. Furthermore, the symmetric configuration of a trilayer sample avoids any flexural strain that can reduce the ME response. The shape demagnetization is an another important parameter that affect the ME response[4, 10-13]. In the same way, the ME response is increased using a trilayer configuration because the magnetic field penetration is improved within two separate ferrite layers in comparison with a single ferrite layer of the same total thickness[12]. However, the calculation of demagnetizing factors is more complicate when two (or more) separated magnetic layers interact together, thus it is difficult to predict and optimize the ME response in such a geometry. There are few publications on this subject in the literature[12,13,14], and to the best of our knowledge, only one concerns a demagnetizing factor calculation using an analytical method[12]. In this paper, we propose an alternative method to calculate the demagnetizing factor of two magnetic discs in a parallel configuration. This analytical calculation is based on a superposition principle and it is valid for a wide range of material permeabilities. Moreover, the model is extended to stacked configurations including more than two magnetic layers. The aim of this work is to quantify and optimize the magnetic field penetration within the ferrites layers in order to maximize the ME response of the ME trilayer. Obviously, the field penetration is better when two magnetic layers are thin and far from each other but at the same time, the mechanical coupling between the magnetic and piezoelectric phases is reduced. So in addition, we have taken into account a mechanical coupling that vary according to the volume ratio of the two phases. We have studied the ME response of ME trilayers theoretically and experimentally in terms of demagnetizing effect and mechanical coupling, and we have investigated the influence of geometrical parameters (thicknesses of PZT and ferrite layers) in order to reach the optimum ME performance of a layered ME composite.

## II. Theoretical basis.

Let us consider a trilayer ME sample made by sandwiching a PZT disc between two ferrite discs, where the direction (3) ($z$ axis) corresponds to the cylindrical axis of symmetry. Usually, the ME effect is measured by applying a small external AC field $H_1^a$ (in direction (1)) superimposed to an external DC bias field. The transversal coupling coefficient $\alpha_{31}$ is then obtained by measuring the induced electrical field $E_3$ in the direction (3). According to this method, the theoretical coupling coefficient (transversal mode) is given by[4]:

$$\alpha_{31} = \frac{E_3}{H_1^a} = -\frac{\eta d_{31}^e}{\varepsilon_{33}^T\left[s_{11}^E + s_{12}^E + \eta\gamma(s_{11}^H + s_{12}^H)\right] - 2(d_{31}^e)^2} \times \frac{(d_{11}^m + d_{12}^m)}{1 + N\chi} \qquad (1)$$

where, for the piezoelectric material, $s_{11}^E$ and $s_{12}^E$ are zero field compliances, $d_{31}^e$ is the piezoelectric coefficient, and $\varepsilon_{33}^T$ is the zero stress permittivity; and for the magnetic material, $s_{11}^H$ and $s_{12}^H$ are zero field compliances, $d_{12}^m$ and $d_{11}^m$ are intrinsic piezomagnetic coefficients, $\chi$ is zero stress dynamic susceptibility, and $N$ is the radial magnetometric demagnetizing factor. $\gamma = V_p/V_f$ is the volume ratio of PZT with respect to ferrite material. The mechanical coupling factor[4] $\eta = \langle S_1^e \rangle / \langle S_1^m \rangle$ (in average), takes into account the differential strain between the PZT layer (average strain: $\langle S_1^e \rangle$) and the ferrite layers (average strain: $\langle S_1^m \rangle$). The ME response curve is mainly shaped by the right hand term in Eq. (1) because the internal DC field $H_{DC}$ sets the value of the intrinsic piezomagnetic coefficients and the dynamic susceptibility. Usually, ME curves are plotted against the external applied DC field $H_{DC}^a$, and due to the demagnetizing effect, the ME curves are shifted along the $H_{DC}^a$ axis because the link between the internal and external DC field is:

$$\vec{H}_{DC} = \frac{1}{1+N.\chi_{DC}} \vec{H}_{DC}^a \tag{2}$$

where $\chi_{DC}$ is the static susceptibility. It must be noted that the radial magnetometric demagnetizing factor $N$ is for two parallel ferrite layers configuration.

The left hand term in Eq. (1) is mainly dependent upon the mechanical properties of the PZT and the ferrite materials, and the mechanical structure of the ME composite. In case of a trilayer sample, the PZT disc is stressed on both faces. The propagation of the longitudinal strain from the two PZT/ferrite interfaces to inner PZT layer depends on the thickness to diameter ratio of the ME sample, and the relative thicknesses of the PZT and ferrite layers. Consequently, the mechanical coupling factor $\eta$ is affected by these parameters. Summarizing, the ME response is affected by the dimensions (thicknesses and diameter) of the layers in the following way: (i) a strong AC field penetration, and consequently a high level ME response are obtained for layer configurations producing low demagnetizing effects; (ii) the demagnetizing factor $N$ sets the optimal working point $H_{DC}^a$ for which the ME response is maximum for a given sample; (iii) the mechanical coupling factor is reduced when the strain distribution is strongly non-uniform across the ME sample. In the next parts of the paper, we will quantify the influences of the geometries (layer sizes) on the ME responses.

## III. Experimental aspect.

### A. ME samples fabrication

Trilayer ME samples were made by bonding a commercial PZT disc with two ferrite discs of the same composition (($Ni_{0.973}Co_{0.027})_{0.875}Zn_{0.125}Fe_2O_4$) chosen for its high piezomagnetic properties[4]. The ferrite material was made by a reactive Spark Plasma Sintering (SPS) technique (see ref. 4). After the SPS stage, all ferrite discs (2 mm in thickness and 10 mm in diameter) were annealed in air at 1000 °C during 1 hour for full re-oxidation. Then the ferrite discs were sliced by means of diamond saw (Struers Secotom-10) and grinded (using silicon carbide papers) to reduce their thickness to appropriate values between 0.165 mm and 1 mm. In the same way, PZT discs (Ferroperm, Pz27, poled along the thickness) with 1 mm or 0.5 mm in thickness and 10 mm in diameter were machined and grinded to reduce their thicknesses from 0.75 mm to 0.25 mm. Lastly, two ferrite discs with the same thickness were pasted (using conductive silver epoxy Epotek E4110) on each faces of the PZT disc. Two sets of samples were prepared. First, ME samples with different thicknesses (0.75, 1.05, 1.5, 2.25, and 3 mm) but with the same PZT/ferrite volume ratio ($\gamma = 0.5$), and secondly, ME samples with the same total thickness (1.5 mm) but with different PZT/ferrite volume ratio ($\gamma = 0.2, 0.5, 1, 2,$ and $3.5$). This relatively simple method of ME samples fabrication process allows however to obtain very accurate characterization results. Indeed, all the PZT discs were poled (with the same optimal electric field) before the fabrication of the ME samples, so, all the ME samples have optimal and reproducible piezoelectric properties. On the other hand, several authors[6,7,9] have already conducted studies on co-sintering PZT/ferrites multilayer. Moreover, in those cases, the co-sintering stage occurs before the poling stage of the PZT layers, and thus the poling process is influenced by the ferrite material for the following reason: (i) the electric field is applied through the ferrite layers which have very low relative permittivity (~10) in comparison to the PZT material (~1000-2000); (ii) if the ferrite material has high resistivity (so no electrical current exists), most of the applied field is absorbed by the ferrite layers. Thus, the poling field within the PZT material may be suboptimal and it may depend on the volume ratio of PZT and ferrite. As a consequence, the piezoelectric properties of the PZT layer and thus the ME properties can vary from a ME sample to an another. This problem is avoided when the PZT material is poled before the fabrication of the ME samples.

**B. ME measurement results and discussion**

The experimental transversal ME coefficient $\alpha_{31}$ is derived from the voltage *V* measured across the electrodes of the PZT layer (direction (3)) when a small external AC magnetic field $H_1^a$ (1 mT in our case) is applied in the direction (1): $\alpha_{31} = V/t_p/H_1^a$, where $t_p$ is the thickness of the PZT layer. The measurements are repeated for different external static field $H_{DC}^a$ (applied in direction (1)) defining the working points ($0 < H_{DC}^a < 6 \times 10^4 A/m$). To avoid any resonance phenomena, the AC magnetic field has been kept low frequency (80 Hz).

In a first experiment, ME coefficients were measured on trilayer samples with always the same PZT/ferrite volume ratio: $\gamma = 0.5$. For a given sample, each layers have the same thickness *t*, and consequently, the total thickness of a sample is *3t*. Experiments were conducted on samples with total thickness between 0.75 mm and 3 mm, and the results are given in Fig. 1. The increase in the sample thickness (for a given PZT/ferrite volume ratio) leads to a decrease in the ME peak amplitude and in a upshift in the peaks positions. These effects are due to the increase of the demagnetizing factor. In fact, the demagnetizing effect reduces the AC field penetration in a ratio $1/(1 + N.\chi)$ and the amplitude of the ME peak is affected accordingly. Furthermore, the DC field penetration is also diminished in a ratio $1/(1 + N.\chi_{DC})$ and thus, the maximums of the intrinsic piezomagnetic coefficients $d_{11}^m$ and $d_{12}^m$ are shifted (by the factor $1 + N.\chi_{DC}$) to higher external DC fields. It must be noted that there is a magnetic coupling between the two ferrite layers and in this case, the demagnetizing factor *N* is higher than the one obtained for a single ferrite layer with the same dimensions. In Fig. 2, the measured ME peak coefficients (circle symbols) and the measured $1/(H_{DC}^a)_{max}$ function (square symbols) are plotted versus the thickness *t*. For better comparisons, the amplitude of the curves were normalized (with respect to the value given by the thicker ME sample). It appears that the two curves are similar, exhibiting the same behavior concerning the demagnetizing effect. It means that in our case, the static permeability $\chi_{DC}$ and the reversible permeability $\chi$ may have a similar value. Using a Vibrating Sample Magnetometer (VSM) technique, the static permeability $\chi_{DC} = M/H$ and the differential permeability $\chi_{diff} = dM/dH$ were measured on a spherical sample of the ferrite material. At the internal field for which the ME coefficients are maximum we obtain: $\chi_{DC} = 84$ and $\chi \sim \chi_{diff} = 94$, which agrees with the previous assumption.

The demagnetizing effect is not the only parameter affecting the ME coefficient $\alpha_{31}$. In Eq. (1), it can be seen that the left hand term is a function of the mechanical coupling $\eta$. We can suppose that this mechanical coupling is improved when the thickness to diameter ratio *t/d* is decreased, thus improving the ME coefficient. On the other hand, as a basic principle, the intrinsic piezomagnetic coefficients are assumed to be unaffected by the mechanical coupling and then, free of the influence of the ratio *t/d*. Thus, $(H_{DC}^a)_{max}$, the field at maximum ME coefficients is free from such an influence. Nevertheless, the two curves plotted in Fig. 2 match well and the $1/(H_{DC}^a)_{max}$ curve is known to be unaffected by the ratio *t/d*. As a consequence, the ME coefficient $\alpha_{31}$, and then the mechanical coupling $\eta$, is free from such an influence. So, another important finding from the previous measurements is that the mechanical coupling $\eta$ seems to be independent of the thickness to diameter ratio *t/d* (in the range $0.025 \leq t/d \leq 0.1$).

The voltage gain, $V = \alpha_{31}.t_p$, is an important parameter for a ME sample used in a real application (a current sensor for example). In our case, when subjected to a 1mT external AC field, the 0.75 mm thick ME sample produces 0.19 V and the 3 mm thick ME sample produces 0.28 V. So when the thickness $t_p$ of the piezoelectric layer is increased by a factor 4, the voltage is increased by a factor 1.5 only. The demagnetizing effect explains this discrepancy: the increase in the ferrite layers thicknesses diminishes the magnetic field penetration. So, increasing the piezoelectric thickness and at the same time decreasing the ferrite layers thicknesses is the way to obtain high voltage gains in trilayer ME samples. To verify this point, trilayer ME samples were fabricated, all with the same total thickness (1.5 mm), but with various PZT/ferrite volume ratio ($\gamma = 0.2, 0.5, 1, 2$, and 3.5). The measured ME voltages are plotted in Fig. 3 for the ME samples subjected to an external AC field of 1mT. It appears that the ME peak voltages increases continuously with the increase of the volume ratio $\gamma$. The maximum voltage (0.47 V) is obtained for the thicker PZT layer ($t_p = 1.17\ mm$). This is due to a combination of two causes. First, for a given electric field *E*, the voltage is proportional to the thickness $t_p$ because $V = E \times t_p$, so increasing $t_p$ will increase *V*. Secondly, the demagnetizing factor is low because the ferrite layers are thin ($t_f = 0.16\ mm$ each) and they are far from each other (distance: $t_p = 1.17\ mm$), so their magnetic interactions are weak. On the other hand, this phenomenon is a little bit counter balanced by a weaker mechanical coupling coefficient $\eta$ at high PZT/ferrite volume ratio $\gamma$. From those experiments, it would appear that the demagnetizing effect is one of the most important phenomenon that influences the ME voltages and the ME coefficients.

## IV. Analytical modeling.

### A. Calculation of magnetometric demagnetizing factors for two parallel ferrite discs

The calculation of the magnetometric demagnetizing factor of a single disc is not trivial. But Chen *et al.* have published several papers on the subject where useful tables of demagnetizing factors for a single disc are given[15,16]. From those data, demagnetizing factors of two ferrites discs in parallel configurations can be derived. The method of calculation is based on a one-dimension superposition method. The calculation is restricted to small thickness $t$ to diameter $d$ ratio (in practice: $t/d \leq 10$ ). In this case, the magnetic materials are assumed (approximately) homogeneously magnetized even for magnetization states far from the saturation.

Let us consider two ferrite discs with the same diameter $d$ and thickness $t$ in a parallel configuration. For simplicity, the distance between the two discs is $t$. As an approximation, we suppose that the two ferrite discs are homogeneously magnetized with the same value $\vec{M}$, parallel to the direction $\vec{x}$ (which is the consequence of an external magnetic field applied in the direction $\vec{x}$). The magnetic structure can be divided into 3 cells: the cells (1) and (3) are the bottom and top ferrite discs, respectively, and the cell (2) is the vacuum between the two discs where $\vec{M} = \vec{0}$ (see Fig. 4). For example, the bottom magnetized ferrite (cell (1)) creates a magnetic field in all the space. This magnetic field inside the cell (1) is called the demagnetizing field whereas the magnetic field created outside the cell (1) is usually called the interaction field. Here, we have chosen to name all those created fields (inside and outside), "dipolar field" because they have the same origin. The total dipolar field $\vec{H}_1^d$ within the bottom ferrite disc (cell (1)) is the sum of two contribution (in average):

$$\vec{H}_1^d = \vec{H}_{1,1}^d + \vec{H}_{3,1}^d \qquad (3)$$

where $\vec{H}_{1,1}^d$ is the dipolar field created by the magnetized cell (1) and acting on it, and $\vec{H}_{3,1}^d$ is the dipolar field created by the magnetized cell (3) and acting on the cell (1). In the same way, the total dipolar field $\vec{H}_3^d$ within the top ferrite disc (cell (3)) is:

$$\vec{H}_3^d = \vec{H}_{3,3}^d + \vec{H}_{1,3}^d \qquad (4)$$

where $\vec{H}_{3,3}^d$ is the dipolar field created by the magnetized cell (3) and acting on it, and $\vec{H}_{1,3}^d$ is the dipolar field created by the cell (3) and acting on the cell (1). Due to symmetry of the problem, $\vec{H}_1^d = \vec{H}_3^d$ and so $\vec{H}_{1,1}^d = \vec{H}_{3,3}^d$ and $\vec{H}_{3,1}^d = \vec{H}_{1,3}^d$.

The global demagnetizing factor $N$ of the two ferrite discs interacting in a parallel configuration (structure given in Fig. 4) can be defined as:

$$\vec{H}_1^d = \vec{H}_3^d = \vec{H}_{3,3}^d + \vec{H}_{1,3}^d = -N.\vec{M} \qquad (5)$$

where $\vec{H}_{3,3}^d$ is the dipolar field of a single magnetized disc. In this case, some tables of calculated demagnetizing coefficient are available in the literature[15,16] and $\vec{H}_{3,3}^d$ can be simply determined from:

$$\vec{H}_{3,3}^d = -N_{t,d}.\vec{M} \qquad (6)$$

where $N_{t,d}$ is the demagnetizing coefficient of a single disc with thickness $t$ and diameter $d$. On the other hand, the determination of $\vec{H}_{1,3}^d$ is not as direct as $\vec{H}_{3,3}^d$. Nevertheless, using a superposition method, the demagnetizing factor $N$ of a magnetic structure consisting in two parallel disc can be derived from a sum of demagnetizing factors of single discs with different thicknesses.

Consider now a single ferrite disc with a diameter $d$ and a thickness $3t$ uniformly magnetized in the direction $\vec{x}$ at the same value $\vec{M}$ (see Fig 5(a)). This disc can be divided into three cells, each with the same thickness $t$. In each cell, we define, $\vec{H}_{i,j}^d$, the dipolar field generated by the cell (i) and acting on the cell (j). So, we obtain the total dipolar field $\vec{H}_1^d$ within the cell (1):

$$\vec{H}_1^d = \vec{H}_{1,1}^d + \vec{H}_{2,1}^d + \vec{H}_{3,1}^d \tag{7}$$

In the same way, the total dipolar field $\vec{H}_2^d$ within the cell (2) is given by:

$$\vec{H}_2^d = \vec{H}_{1,2}^d + \vec{H}_{2,2}^d + \vec{H}_{3,2}^d \tag{8}$$

And $\vec{H}_3^d$ within the cell (3) is given by:

$$\vec{H}_3^d = \vec{H}_{1,3}^d + \vec{H}_{2,3}^d + \vec{H}_{3,3}^d \tag{9}$$

Due to the geometry and the symmetry of the problem, $\vec{H}_{i,j}^d = \vec{H}_{j,i}^d$ if $i \neq j$ and if $i = j$ and $\vec{H}_{1,1}^d = \vec{H}_{2,2}^d = \vec{H}_{3,3}^d$. Thus, the mean value of the global dipolar field within a single disc with a thickness $3t$ can be written:

$$\langle \vec{H}_{123}^d \rangle = (\vec{H}_1^d + \vec{H}_2^d + \vec{H}_3^d)/3 = \vec{H}_{3,3}^d + \frac{2}{3}\vec{H}_{1,3}^d + \frac{4}{3}\vec{H}_{2,3}^d = -N_{3t,d}.\vec{M} \tag{10}$$

where $N_{3t,d}$ is the demagnetizing factor of a single ferrite disc uniformly magnetized with a thickness $3t$ and a diameter $d$.

For a single ferrite disc with a diameter $d$ and a thickness $2t$ uniformly magnetized and divided into two cells (region (2) and (3)) with the same thickness $t$ (see Fig. 5(b)), the mean value of the global dipolar field is:

$$\langle \vec{H}_{23}^d \rangle = (\vec{H}_{3,3}^d + \vec{H}_{2,3}^d + \vec{H}_{2,2}^d + \vec{H}_{3,2}^d)/2 = \vec{H}_{3,3}^d + \vec{H}_{2,3}^d = -N_{2t,d}.\vec{M} \tag{11}$$

where $N_{2t,d}$ is the demagnetizing factor of a single ferrite disc uniformly magnetized with thickness $2t$ and diameter $d$.

Lastly, considering a single ferrite disc with a diameter $d$ and a thickness $t$ uniformly magnetized (cell (3)) (see Fig. 5(c)), the mean value of the global dipolar field is:

$$\langle \vec{H}_3^d \rangle = \vec{H}_{3,3}^d = -N_{t,d}.\vec{M} \tag{12}$$

where $N_{t,d}$ is the demagnetizing factor of a single ferrite disc uniformly magnetized with thickness $t$ and diameter $d$.

Combining Eqs. (10), (11), (12), leads to:

$$\vec{H}_{1,3}^d = -(\frac{3}{2}N_{3t,d} - 2.N_{2t,d} + \frac{1}{2}N_{t,d}).\vec{M} \tag{13}$$

Combining Eqs. (5), (6), and (13), we obtain the dipolar field within two ferrite discs, each with thickness $t$ and diameter $d$, separated by a distance $t$, and interacting in a parallel configuration:

$$\vec{H}_1^d = \vec{H}_3^d = \vec{H}_{3,3}^d + \vec{H}_{1,3}^d = -(\tfrac{3}{2}N_{3t,d} - 2.N_{2t,d} + \tfrac{3}{2}N_{t,d}).\vec{M} = -N.\vec{M} \tag{14}$$

where $N$ is the magnetometric demagnetizing factor for such a magnetic structure:

$$N = \tfrac{3}{2}N_{3t,d} - 2.N_{2t,d} + \tfrac{3}{2}N_{t,d} \tag{15}$$

When the two parallel ferrite discs are spaced with a distance $e$ different from the thickness $t$ of a disc, using the previous method, the calculation of the magnetometric demagnetizing factor leads to:

$$N = N_{t,d} - \left(1 + \tfrac{e}{t}\right).N_{e+t,d} + \left(1 + \tfrac{e}{2t}\right).N_{e+2t,d} + \tfrac{e}{2t}.N_{e,d} \tag{16}$$

where $N_{e+t,d}$, and $N_{e+2t,d}$ are demagnetizing factors for single ferrite discs with thicknesses $(e + t)$ and $(e + 2.t)$ respectively, and diameter $d$. Eq. (16) is similar to the one calculated by Liverts et al.[12] for two parallel rectangular ferromagnetic prisms. A useful formula for the calculation of the radial demagnetizing factor $N_{t,d}$ of a single disc is given in Appendix A.

## B. Estimation of the mechanical coupling behavior

In a stacked ME sample, the mechanical coupling is due to the shear stress that propagates from the ferrite layers to the PZT layer through the interfaces. The induced strain field results from the equilibrium between shear and extensional stresses in each layers. The ME response is obtained when a small alternative strain field (produced by the alternative magnetic field) is superimposed with a DC strain field (produced by the bias magnetic field). Consequently, the strain distribution is inhomogeneous in a ME sample and the exact alternative strain field can be accurately predicted only by numerical methods[17] (Finite Element Method for example). Nevertheless, in some special cases, the mechanical coupling could be estimated as following. When the PZT layer thickness $t_p$ is very small compared to the total thickness of the ferrite layers $t_f$, a far-field strain is obtained in the PZT layer (because $t_p \ll d$) and the ferrite layers are almost mechanically free. Thus, the strain fields are almost homogeneous and equal in each layer and the relative differential strain is close to zero:

$$\frac{\langle S_f \rangle - \langle S_p \rangle}{\langle S_f \rangle} \sim 0 \tag{17}$$

On the other hand, when the ferrite layers thickness is very small compared to the PZT one ($t_f \ll t_p$), the strain field is almost homogeneous in the ferrite layers (far-field approximation) and merges with the PZT/ferrite interfaces strain $\langle S_{p,f} \rangle$. The problem is now resumed to that of a PZT disc stressed on its both faces. In this case, the relative differential strain have a finite value $a$, ($0 < a < 1$) which depends on the mechanical properties and dimensions of the PZT layer:

$$\frac{\langle S_f \rangle - \langle S_p \rangle}{\langle S_f \rangle} \sim \frac{\langle S_{p,f} \rangle - \langle S_p \rangle}{\langle S_f \rangle} \sim a \tag{18}$$

These two extreme situations suggest that $(\langle S_f \rangle - \langle S_p \rangle)/\langle S_f \rangle$ is a function of $t_p/(t_p + t_f)$ rather than $t_p/t_f$, because when $t_f \ll t_p$, the ratio $t_p/t_f$ tends towards infinity, which contradicts Eq. (18). Since the thickness to diameter ratio of each layer of our ME samples are less than 10 %, we can assume a first order approximation in the mechanical coupling behavior, and then:

$$\frac{\langle S_f \rangle - \langle S_p \rangle}{\langle S_f \rangle} \sim a \cdot \frac{t_p}{t_p + t_f} \qquad (19)$$

in the range $0 < t_p/(t_p + t_f) < 1$. From the previous equation, we deduce an approximated mechanical coupling factor $\eta$:

$$\eta = \frac{\langle S_p \rangle}{\langle S_f \rangle} = -\frac{\langle S_f \rangle - \langle S_p \rangle}{\langle S_f \rangle} + 1 \sim -a \cdot \frac{t_p}{t_p + t_f} + 1 \qquad (20)$$

The elastic bonding layers (silver epoxy in our case) absorb a part of the stress at the PZT/ferrite interfaces[13] and the result is a downshift of the mechanical coupling factor:

$$\eta \sim -a \cdot \frac{t_p}{t_p + t_f} + b \qquad (21)$$

where $b < 1$ is the coupling factor of the bonding layers (obtained when $t_p \ll t_f$). The values of *a* and *b* in Eq. (21) are unknown and are therefore fitting parameters.

**C. Application to an analytical modeling of the ME effect in trilayer ME samples**

To calculate the ME response of trilayer samples, the theoretical demagnetizing factor *N* of two parallel ferrite discs must be derived from the previous theory. First, radial magnetometric demagnetizing factors for a single ferrite disc were interpolated ($\chi = 94$) from data published by Chen *et al.*[15,16] (circle symbols in Fig. 6). These results were fitted using a polynomial function of degree 3 (for $0.01 \leq t/d \leq 0.3$):

$$N_{t,d} \approx A + B \times (t/d) + C \times (t/d)^2 + D \times (t/d)^3 \qquad (22)$$

where the polynomial coefficients are: *A*=0.0027, *B*=1.014, *C*=-2.087, *D*=2.313. The fitted result is plotted in dashed line in Fig. 6. Then, an analytical approximation of the demagnetizing factor *N* of two ferrite discs in parallel configuration (with thickness *t* each) separated by a distance *t* is calculated using Eqs. (15). To verify our analytical calculation for two parallel discs, (solid line in Fig. 6), we have solved the same magnetic problem by a numerical method based on a Finite Element Method (FEM) software (ANSYS Maxwell). The numerical results plotted in Fig. 6 (square symbols) validate the analytical method of demagnetizing calculation developed in this paper. To verify the relationship between the ME response and the demagnetizing effect, the theoretical field reduction ratios $1/(1 + N\chi)$ were calculated, normalized, and plotted in Fig. 2 (triangle symbols) for the five ME samples. There is a good agreement between the theory and the experiment that confirm the influence of the demagnetizing effect, except for the thinnest sample ($t = 0.25\ mm$). In this case, the experimental result is 18 % over the theoretical one. It suggests that the magnetic properties are different for this sample: we may suppose that the AC susceptibility is lower than 94 and then, the field penetration is improved, leading to a stronger ME response. In this study, since the PZT/ferrite volume ratio $\gamma$ is maintained constant in the ME samples, the mechanical coupling factor $\eta$ seems to have a constant value ($\eta \sim 0.7$). On the other hand, the ME voltages presented in Fig. 3 are obtained for ME samples whose $\gamma$ ratios vary in a large proportion ($0.2 \leq \gamma \leq 3.5$), involving a large variation in the mechanical coupling factor $\eta$. The peak ME voltages of these samples were theoretically calculated using Eq. (1), by including both the demagnetizing effect (using Eq. (16)) and the mechanical coupling. Materials properties used for the calculation are summarized in Table I. It must be noted that $d_{11}^m$ is the intrinsic piezomagnetic coefficient and $d_{12}^m \sim -d_{11}^m/2$ for polycrystalline Ni-Co-Zn ferrites. First, we have plotted the theoretical peak ME voltages (dotted and dashed lines in Fig. 7), taking into account constant mechanical coupling factors $\eta = 1$ and $\eta = 0.71$ repectivly (see Ref. 4 for the choice of this value). We see that the theory do not match with the experiments when the value of the mechanical coupling factor is assumed constant. Then, according to Eq. (21), we have introduced a factor $\eta$ that decreases linearly against the increase of the PZT volume ratio $V_p/(V_p + V_f)$. The obtained theoretical ME voltage, plotted in solid line in Fig. 7 shows an improved agreement between theory and measurements, especially for the ME samples with a thick PZT layer. The linear curve $\eta$ that permits to fit the experimental ME voltages is plotted in Fig. 8 in dashed line. In the $V_p/(V_p + V_f) \sim 0$ extrapolated area, the mechanical coupling factor is lower than 100%, which means that the strains are not perfectly transmitted through the glue layers, and a fraction (20%) is absorbed. The analytical demagnetizing factor $N$ used in the ME voltage calculation and plotted in Fig. 8 is nearly linear as confirmed by the FEM software calculations (square symbols). Since all samples have the same total thickness (1.5 mm), the ME coefficient normalized with respect to the total thickness (1.5 mm) is given on the right vertical axis in Fig. 7. Experimentally, the ME coefficient reach $\alpha_{31} = 0.4\, V/A$ for the optimal ME sample ($\gamma = 0.35$). Nevertheless, even for this optimized sample, due to the demagnetizing effect, the internal magnetic field reaches only 30 % of the external applied field (see Fig. 9). This low value is due to the high dynamic magnetic susceptibility ($\chi = 94$) that appears in the term $1/(1 + N\chi)$.

**D. Numerical calculation of the mechanical coupling factor.**

In the previous part, it was shown that the coupling factor $\eta$ is strongly affected by the epoxy bonding layers, and even in the best case, more than 20% of the strain is absorbed by those layers. In fact, there are mechanical properties mismatches between the ceramic materials (Young modulus: $E = 59\, GPa$, for the PZT and $E = 154\, GPa$ for the ferrite) and an epoxy resin ($E = 3 - 10\, GPa$, depending on material references), leading to a strong shear strain field within the bonding layers[13]. Some authors[18] have shown that the mechanical properties of the epoxy layers have high influence on the ME response of a ME device. Furthermore, a coupling coefficient[18] must be introduced to model an interface detachment between ceramic layers (PZT and ferrite) and the epoxy layers. In order to evaluate the influence of the bonding layers on the mechanical coupling factor $\eta$, the strain field within the magnetoelectric structure was modeled using a 3-dimensional Finite Element Method (ANSYS software). The simulated mechanical structure consists in two ferrite discs bonded by two epoxy layers on both face of the PZT disc. The thickness of each epoxy layer (almost 30µm) was measured by means of an optical microscope. This relatively high thickness can be explained by the size of the silver particles constituting the conductive filler. The manufacturer (Epoteck) indicates particle sizes lower than 48µm, which can explains the thickness of the bounding layers. The mechanical properties were experimentally obtained from ultrasonic velocity measurements in a sample of Epotek E4110 (pulse-echo method[4]). The results are: Young modulus, $E = 8\ GPa$, and Poisson ratio, $\nu = 0.35$ (estimation).

For a given AC magnetic field excitation, the strain field was calculated in each layers (using FEM method) and lastly, the theoretical mean strain ratio in the PZT and ferrite layer, $\eta = \langle S_1^e \rangle / \langle S_1^m \rangle$, was obtained. The FEM calculations were done for each of the five samples of the study. These theoretical results are given in Fig. 10, in comparison with the experimental coupling factor used to fit the data (solid line). First, we studied a structure where the epoxy layers are perfectly mechanically coupled to the PZT and ferrite layers. This means that there is no sliding at the PZT/epoxy and ferrite/epoxy interfaces, or in other words, the coefficient of friction is $k = 1$ at the interface (square symbols in Fig. 10). It is seen that the behavior is almost linear with the PZT volume ratio (as predicted by Eq. 21), with a slope close to the experimental curve (solid line), but with a bias overestimation. To overcome this systematic error, the same structure was simulated, but we have introduced a coefficient of friction modeling a sliding produced by local interface detachments or cracks within the epoxy layers (square symbols). The value $k = 0.25$ was chosen to fit the experimental coupling factor $\eta$. Lastly, for comparison, we have studied the case of an assumed structure, where the ferrite discs are perfectly clamped ($k = 1$, no sliding) on both faces of the PZT disc without intermediate layers of epoxy resin (triangle symbols). In this case, the coupling factor reflects only the shear strain effect within the PZT and ferrite layers. This mechanical study shows that the differential strain between the PZT and ferrite layers is mainly due to the shear strain within the epoxy layers and the sliding at the PZT/epoxy and ferrite/epoxy interfaces. The shear strain along the thickness of the PZT layer plays a secondary role.

**E. Analytical calculation of demagnetizing effect for multilayered ME samples**

In the previous part, we have demonstrated that a low demagnetizing effect, so a high ME voltage is obtained in a ME trilayer when the two magnetic layers are thin and far from each other. By increasing the number of layers ($n > 3$) at a given PZT/ferrite volume ratio and at a given total thickness, we may assume a reduction of the global demagnetizing effect because the magnetic layers are thinner. But in the other hand, in this case, the magnetic layers are closer to each other and this tends to counterbalance the previous effect, and it is difficult to say which phenomenon dominates. To answer this question, using the calculation method presented in section IV. A. (with the same restrictions), we have derived an analytical formula giving the global demagnetizing factor of a multilayered ME sample. This formula is an extrapolation of Eq. (16) which gives the demagnetizing factor of two magnetic discs (Fig. 4). In this simple structure, the demagnetizing factor of a layer is the sum of two contributions. The first one, $N_{t,d}$ is the influence of a given layer on itself. The second one, that we call $\mathbb{N}_{t,e,d}$, is the influence from the other layer (with thickness *t* and diameter *d*) situated at a distance *e*, where:

$$\mathbb{N}_{t,e,d} = -\left(1+\frac{e}{t}\right).N_{e+t,d} + \left(1+\frac{e}{2t}\right).N_{e+2t,d} + \frac{e}{2t}.N_{e,d} \qquad (23)$$

Consider now a stack of *n* alternated ferrite layers (thickness *t* each) and PZT layers (thickness *e* each), with a ferrite layer at the top and bottom of the stack, so that *n* is an odd number (see Fig 11). Then, the radial demagnetizing factor $N_{t,d}^p$ of a ferrite layer numbered *p* (odd number between 1 and *n*), is the sum of the $N_{t,d}$ term (the influence of the layer *p* on itself) and all the $\mathbb{N}_{t,x,d}$ terms produced by each of the other layers distant of $x$, where $x = e,\ 2e+t,\ 3e+2t, \ldots, \left(\frac{n-p}{2}\right)e + \left(\frac{n-p}{2}-1\right)t$, for the magnetic layers above the layer *p*, and where $x = e,\ 2e+t,\ 3e+2t, \ldots, \left(\frac{p-1}{2}\right)e + \left(\frac{p-1}{2}-1\right)t$, for the magnetic layers below the layer *p*. Then, the radial demagnetizing factor $N_{t,d}^p$ of a ferrite layer numbered *p* can be written as:

$$N_{t,d}^p = \sum_{k=1}^{(p-1)/2} \mathbb{N}_{t,(ke+(k-1)t),d} + \sum_{k=1}^{(n-p)/2} \mathbb{N}_{t,(ke+(k-1)t),d} + N_{t,d} \qquad (24)$$

Replacing the $\mathbb{N}_{t,(ke+(k-1)t),d}$ term in Eq. (24) by its expression given in Eq. (23), we obtain:

$$N_{t,d}^p = \sum_{k=1}^{\frac{p-1}{2}} \begin{bmatrix} \frac{ke+(k+1)t}{2t} N_{ke+(k+1)t,d} \\ -\frac{k(t+e)}{t} N_{k(t+e),d} \\ +\frac{ke+(k-1)t}{2t} N_{ke+(k-1)t,d} \end{bmatrix}$$

$$+ \sum_{k=1}^{\frac{n-p}{2}} \begin{bmatrix} \frac{ke+(k+1)t}{2t} N_{ke+(k+1)t,d} \\ -\frac{k(t+e)}{t} N_{k(t+e),d} \\ +\frac{ke+(k-1)t}{2t} N_{ke+(k-1)t,d} \end{bmatrix}$$

$$+ N_{t,d} \qquad (25)$$

where *t* is the thickness of each ferrite layer, and *e* is the thickness of each PZT layer; $N_{x,d}$ is the radial demagnetizing factor of a single magnetic disc with diameter *d* and thickness *x*. The global demagnetizing factor $\langle N \rangle$ of the stack is deduced from the average over all magnetic layers:

$$\langle N \rangle = \frac{1}{(n+1)/2} \sum_{k=0}^{(n-1)/2} N_{t,d}^{p=2k+1} \qquad (26)$$

where $(n+1)/2$ is the quantity of magnetic layers.

Using Eq. (25) and (26), the global demagnetizing factor was calculated for a number of layers between 3 and 21 at different values of PZT/ferrite volume ratio $V_p/V_f$ for a ME sample of 1.5 mm total thickness and 10 mm diameter. The theoretical calculations are plotted in Fig. 12. The result shows that, for a given PZT/ferrite volume ratio, the global demagnetizing factor is almost independent of the number of magnetic layers in the ME stack. So, in theory, from a demagnetizing effect point of view, the ME voltage cannot be enhanced by increasing the number of layers. On the other hand, in terms of mechanical coupling effect, we can expect a better strain uniformity for structures comprising large numbers of layers, which means that the mechanical coupling factor approaches its maximum value (we may assume $\eta \sim 0.8$ at best). In this later case, the ME voltage can be improved by 20 %. Fig. 13 shows the theoretical profile of the demagnetizing factor through the thickness for three different ME stacks ($n = 7, 11,$ and $21$) at a given PZT/ferrite volume ratio ($V_p/V_f = 3.5$). Obviously, demagnetizing factors are higher for magnetic layers close to the centre of the stack (where the influences of all the magnetic layers are stronger), and lower for the external layers. However, the inhomogeneity is lower than 17% whatever the number of layers, which validates the use of a global demagnetizing factor $\langle N \rangle$ averaged over the whole stack.

## V. Conclusion.

We have proposed an analytical model suitable for the calculation of the demagnetizing field in ME trilayers where two magnetic discs interact together in a parallel configuration. This analytical model is based on a superposition method, involving the demagnetizing factors of single magnetic discs. A Finite Element Method was used to solve numerically the magnetic problem, permitting to validate the analytical approach. The analytical demagnetizing factors were introduced into a model predicting the ME voltage coefficient. It appears that the theoretical calculations fit well the experimental ME responses when a mechanical coupling factor depending on the PZT/ferrite volume ratio is introduced. This work has revealed that a maximum ME voltage is reached for ME samples with high PZT volume ratios. In this case, the low mechanical coupling between the PZT and ferrite layers is counter balanced by a better magnetic field penetration because the two ferrite layers are thin and are relatively far from each other. This implies that the diameter and the thickness of the PZT and ferrites layers are the geometrical parameters that affect the ME response through the demagnetizing effect and the mechanical coupling. The superposition method that we used to calculate the demagnetizing factor of a trilayer ME sample has been extended to multilayered ME samples. The important finding is that an increase of the number of layers (at a given sample thickness and PZT/ferrite volume ratio) do not change the magnetic field penetration, and the ME voltage remains unchanged from this point of view. Summarizing, the ME trilayer structure combines the advantages of high ME performance and ease of fabrication.

# Appendix A: magnetometric radial demagnetizing factor calculation for a single disc.

This appendix is an additional part of Sec. IV. A formula for the calculation of the magnetometric radial demagnetizing factor of a single magnetic disc is given below. The validity domain is: susceptibility $\chi$ between 1 and 190, and thickness to diameter ratio t/d between 0.01 and 0.5, which satisfy most of the layered ME samples with cylindrical geometries. This formula is derived from the works of Chen *et al.*[15,16].

$$N_{t,d}(\chi) = (C_{11}\chi^4 + C_{12}\chi^3 + C_{13}\chi^2 + C_{14}\chi + C_{15}) \cdot \left(\frac{t}{d}\right)^4$$

$$+(C_{21}\chi^4 + C_{22}\chi^3 + C_{23}\chi^2 + C_{24}\chi + C_{25}) \cdot \left(\frac{t}{d}\right)^3$$

$$+(C_{31}\chi^4 + C_{32}\chi^3 + C_{33}\chi^2 + C_{34}\chi + C_{35}) \cdot \left(\frac{t}{d}\right)^2$$

$$+(C_{41}\chi^4 + C_{42}\chi^3 + C_{43}\chi^2 + C_{44}\chi + C_{45}) \cdot \left(\frac{t}{d}\right)$$

$$+(C_{51}\chi^4 + C_{52}\chi^3 + C_{53}\chi^2 + C_{54}\chi + C_{55})$$

Where the $C_{ij}$ coefficients are:

$C_{11} = -1.27 \times 10^{-8}$;  $C_{12} = 5.69 \times 10^{-6}$;  $C_{13} = -8.79 \times 10^{-4}$;

$C_{14} = 5.42 \times 10^{-2}$;  $C_{15} = -2.88$;

$C_{21} = 1.41 \times 10^{-8}$;  $C_{22} = -6.29 \times 10^{-6}$;  $C_{23} = 9.67 \times 10^{-4}$;

$C_{24} = -5.89 \times 10^{-2}$;  $C_{25} = 4.15$;

$C_{31} = -5.42 \times 10^{-9}$;  $C_{32} = 2.41 \times 10^{-6}$;  $C_{33} = -3.67 \times 10^{-4}$;

$C_{34} = 2.20 \times 10^{-2}$;  $C_{35} = -2.55$;

$C_{41} = 9.78 \times 10^{-10}$;  $C_{42} = -4.33 \times 10^{-7}$;  $C_{43} = 6.55 \times 10^{-5}$;

$C_{44} = -3.90 \times 10^{-3}$;  $C_{45} = 1.08$;

$C_{51} = 2.57 \times 10^{-11}$;  $C_{52} = -1.23 \times 10^{-8}$;  $C_{53} = 2.10 \times 10^{-6}$;

$C_{54} = -1.55 \times 10^{-4}$;  $C_{55} = 7.00 \times 10^{-3}$;

# References.

|  | $d_{31}^e$ (pC/N) | $d_{11}^m$ (nm/A) | $s_{11}^E$ or $s_{11}^H$ (m²/N) | $s_{12}^E$ or $s_{12}^H$ (m²/N) | $\mu^T$ or $\varepsilon_{33}^T$ (in relative) |
|---|---|---|---|---|---|
| Pz27 | -170 |  | $17 \times 10^{-12}$ | $-6.6 \times 10^{-12}$ | 1800 |
| ferrite |  | -9.5 | $6.47 \times 10^{-12}$ | $-1.84 \times 10^{-12}$ | 95 |

TABLE 1 : Material properties for Pz27 (cited from Ferroperm[19]), and Ni-Co-Zn ferrite (Ref. 4). Note that $d_{11}^m$ is the intrinsic piezomagnetic coefficient.

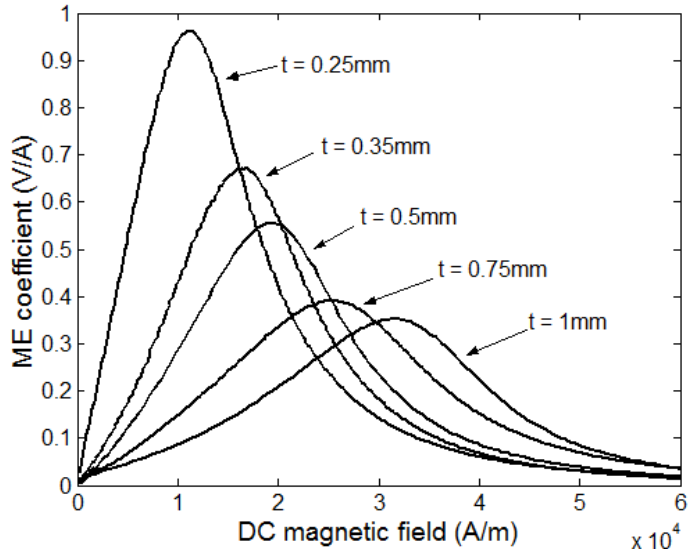

FIG. 1. Magnetoelectric coefficients for trilayer sample with various thicknesses. $t$ is the thickness of a ferrite layer (or a PZT layer). The total thickness of a sample is $3t$.

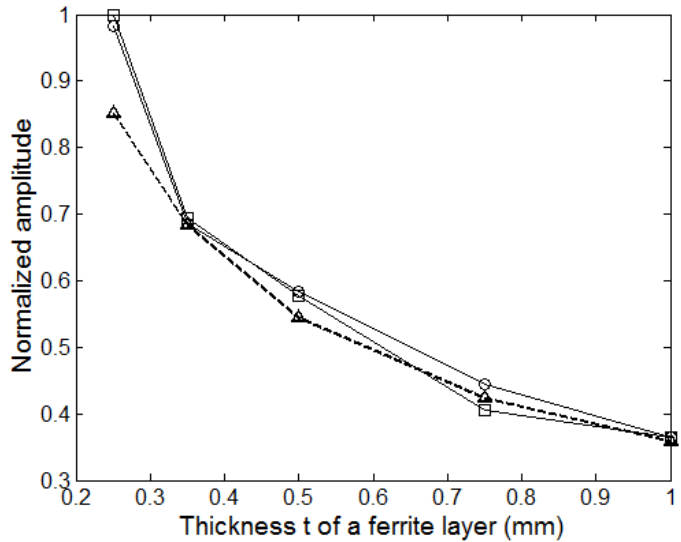

FIG. 2. Circles: normalized measured ME peak coefficients. Squares: normalized measured $1/(H_{DC}^a)_{max}$ function. Triangles: normalized theoretical ME peak coefficients. $t$ is the thickness of a ferrite layer (or a PZT layer). The total thickness of a sample is $3t$.

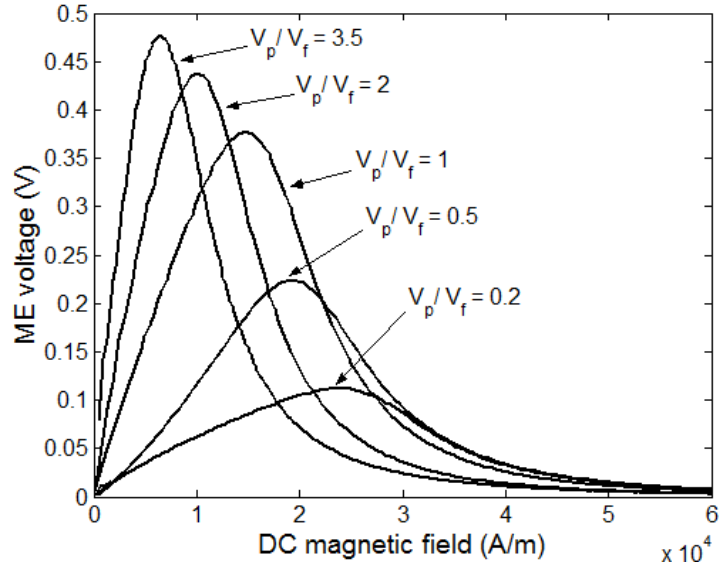

FIG. 3. Magnetoelectric voltages for trilayer samples with various PZT/ferrite volume ratio $\gamma = V_p/V_f$. All the samples have the same total thickness (1.5 mm). The amplitude of the external AC field is 1 mT.

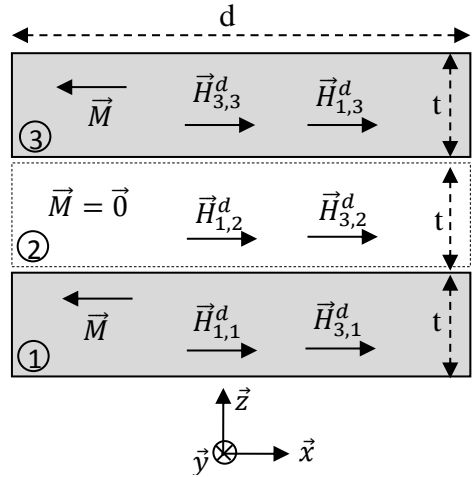

FIG 4. Sketches of a magnetic structure consisting in two ferrite discs uniformly magnetized (cells (1) and (3)) in a parallel configuration separated by a layer of vacuum (cell (2) in dashed line).

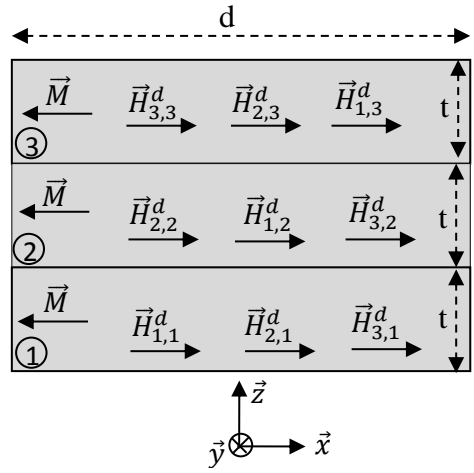

(a)

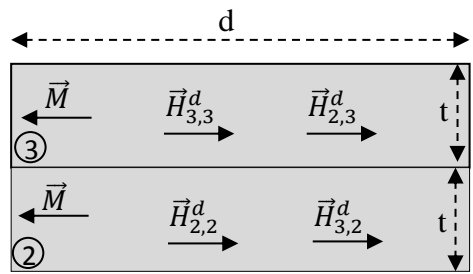

(b)

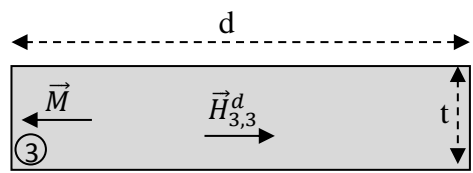

(c)

FIG 5. Sketches of three different magnetic structures; (a) a single magnetic disc divided into three cells with thickness *t* each; (b) a single magnetic disc divided into two cells; (c) a single magnetic disc corresponding to a unique cell.

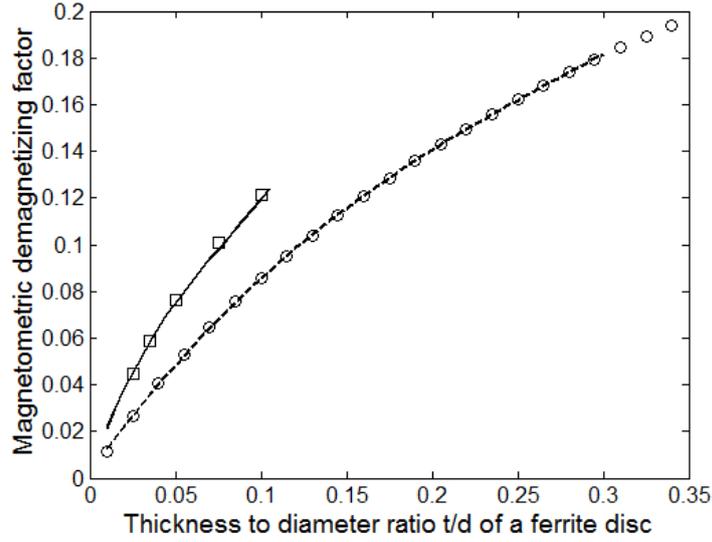

FIG 6. Theoretical magnetometric demagnetizing factor. Circle symbols: single ferrite disc (data derived from Chen *et al.*). Dashed line: polynomial fit. Solid line: demagnetizing factor for two parallel discs (distance: *t*) deduced from data of a single ferrite disc. Square symbols: numerical calculation for two parallel discs. In all cases, $\chi = 94$.

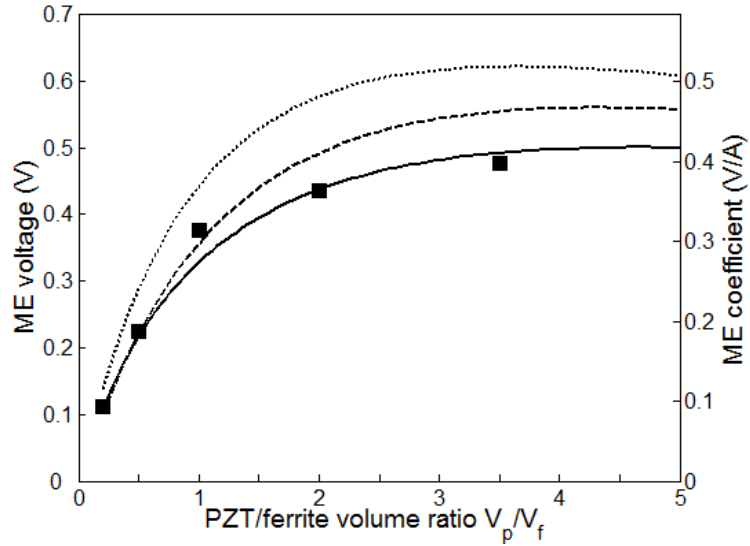

FIG 7. Theoretical ME peak voltages as a function of the PZT/ferrite volume ratio $\gamma$. Dashed and dotted lines: the mechanical coupling factor $\eta$ is maintained constant. Dotted line: $\eta = 1$ and dashed line: $\eta = 0.71$. Solid line: the mechanical coupling factor decrease linearly with the increasing thickness of the PZT layer. Square symbols: experimental peak ME voltages extracted from Fig. 3. The corresponding ME coefficient normalized with respect to the total thickness (1.5 mm) of the ME samples is given on the right vertical axis.

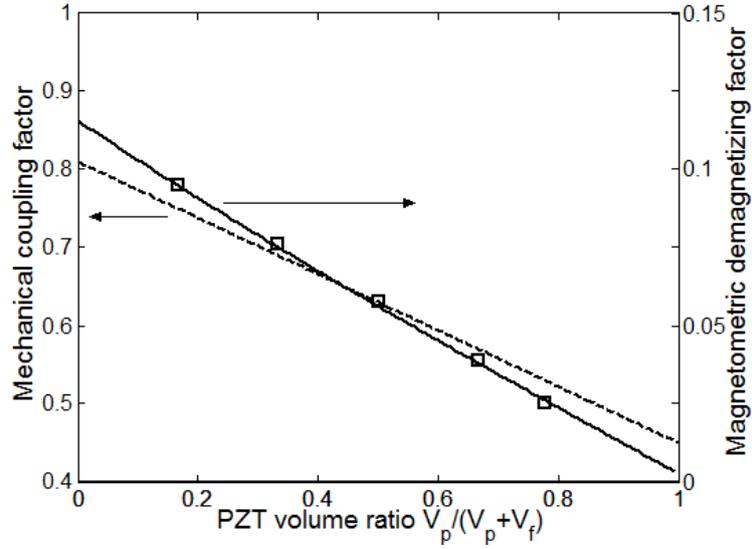

FIG 8. Mechanical coupling factor $\eta$ (dashed line) and magnetometric demagnetizing factor $N$ (solid line) used for the analytical calculation of the ME voltages and plotted against the PZT volume ratio $V_p/(V_p + V_f)$. Square symbols: numerical calculation (FEM software) of demagnetizing factors ($\chi = 94$) for comparison.

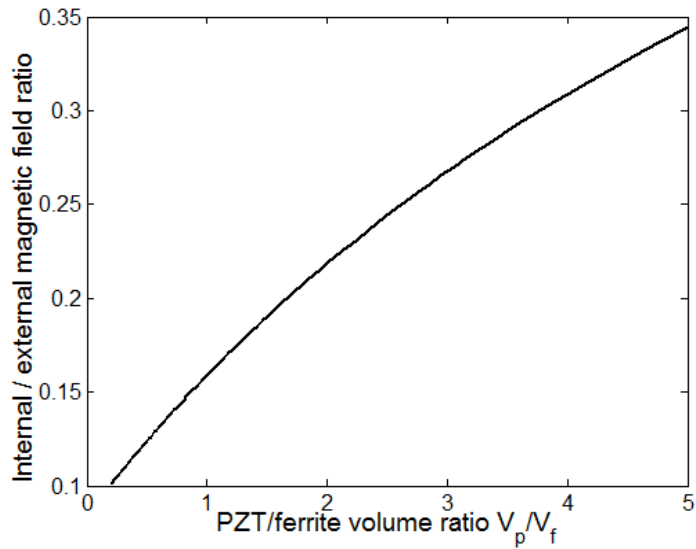

FIG 9. Theoretical ratio between the internal magnetic field and the external field as a function of the PZT/ferrite volume ratio.

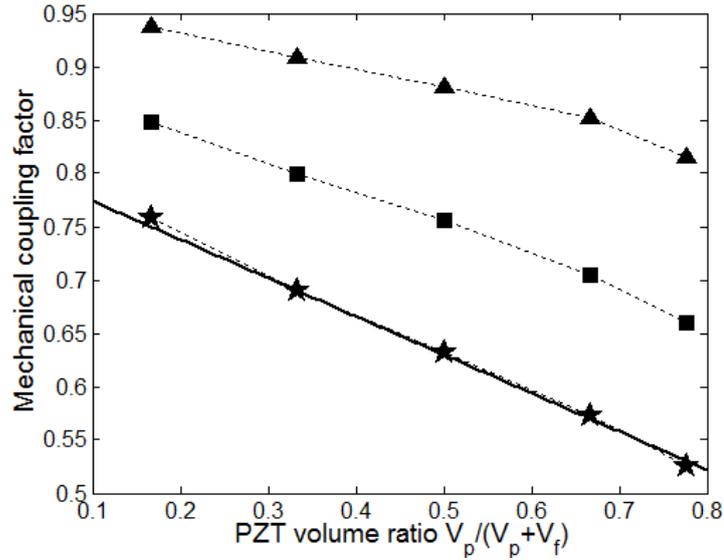

FIG 10. Mechanical coupling factor $\eta$ versus PZT volume ratio. Solid line: linear function that permits to fit the experimental ME voltages. Symbols: FEM calculation results. Square symbols: structure with epoxy layers perfectly coupled ($k = 1$) to the PZT and ferrite layers. Star symbols: structure with sliding at the interfaces ($k = 0.25$). Triangle symbols: structure assuming ferrite and PZT layers perfectly coupled (without intermediate layers). Dotted lines are linear interpolations.

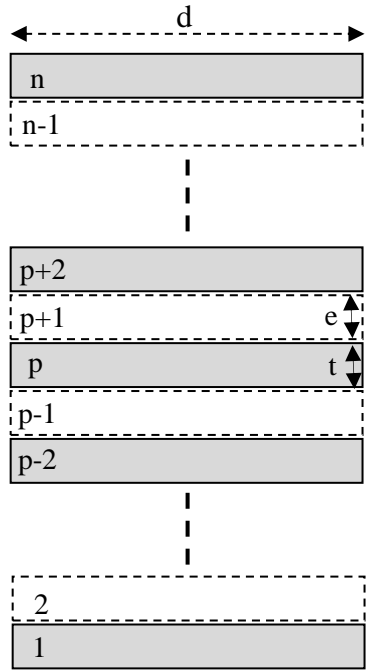

FIG 11. Sketches of a multilayered structure consisting in (n+1)/2 ferrite discs uniformly magnetized (cells in grey) in a parallel configuration separated by (n-1)/2 PZT discs (cells in dashed line).

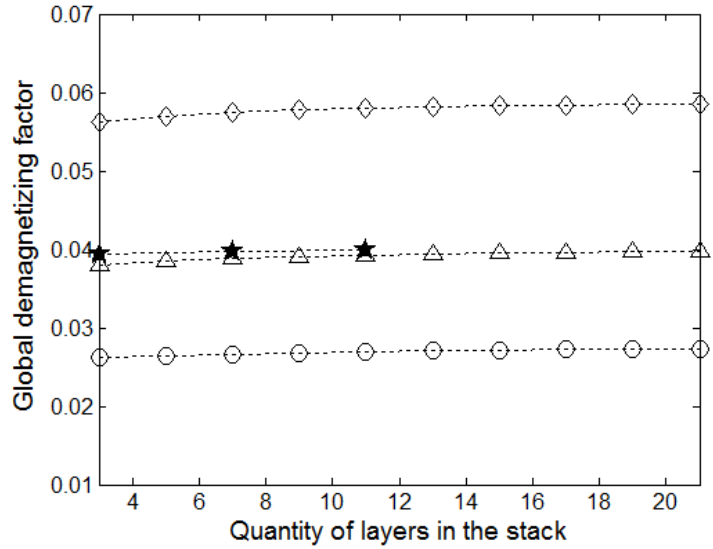

FIG 12. Theoretical global demagnetizing factor, as a function of the quantity of layers in the ME stack, plotted for three different PZT/ferrite volume ratios. Circle symbols: $V_p/V_f = 3.5$; triangle symbols: $V_p/V_f = 2$; diamond symbols: $V_p/V_f = 1$. Star symbols: numerical FEM calculation ($V_p/V_f = 2$) for comparison. In each cases, the total thickness of a ME sample is 1.5 mm and the diameter is 10 mm.

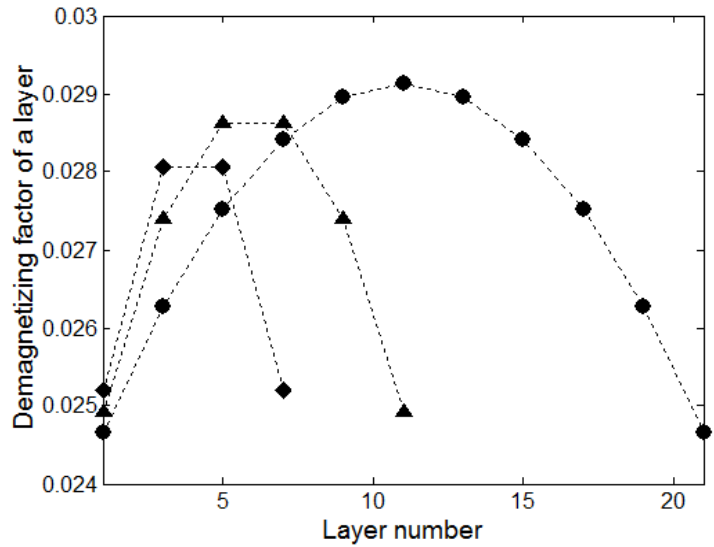

FIG 13. Theoretical profile of the demagnetizing factor within ME stacks for a volume ratio $V_p/V_f = 3.5$. Circle symbols: 21 layers in the stack ; triangle symbols: 11 layers in the stack; diamond symbols: 7 layers in the stack. In each cases, the total thickness of a ME sample is 1.5 mm and the diameter is 10 mm.